\documentclass[conference]{IEEEtran}
\IEEEoverridecommandlockouts
\usepackage{cite}
\usepackage{amsmath,amssymb,amsfonts}
\newcommand{\norm}[1]{\left\lVert#1\right\rVert}
\usepackage{algorithmic}
\usepackage{graphicx}
\usepackage{svg}
\usepackage{textcomp}
\usepackage{tikz}
\usetikzlibrary{plotmarks}
\usepackage{pgfplots}
\usepackage{caption}
\usepackage{subcaption}
\usepackage{lipsum}
\usepackage{siunitx}
\usepackage{mwe}
\usepackage{xcolor}
\def\BibTeX{{\rm B\kern-.05em{\sc i\kern-.025em b}\kern-.08em
		T\kern-.1667em\lower.7ex\hbox{E}\kern-.125emX}}
\usepackage{amsthm}
\usepackage{mathtools}

\begin{document}
	
	\title{Optimizing Energy Efficiency with RSMA: Balancing Low and High QoS Requirements\\
		{}
		\thanks{This work was funded by the German Federal Ministry of Education and Research (BMBF) in the course of the 6GEM research hub under grant number 16KISK037.}
	}

	\author{\IEEEauthorblockA{ Srivardhan Sivadevuni, Kevin Weinberger and Aydin Sezgin}
		\IEEEauthorblockA{
			Ruhr University Bochum, Germany \\
			Email: \{srivardhan.sivadevuni, kevin.weinberger, aydin.sezgin\}@rub.de}
	}
	\maketitle

 \begin{abstract}
     Future wireless systems are expected to deliver significantly higher quality-of-service (QoS) albeit with fewer energy resources for diverse, already existing and also novel wireless applications. The optimal resource allocation for a system in this regard could be investigated by reducing the overall power available at the expense of reduced QoS for the inefficient users. In other words, we maximize the system energy efficiency by achieving power saving through a minimal back-off in terms of QoS. In this paper, we investigate the energy efficiency vs. delivered QoS trade-off for the rate-splitting multiple access (RSMA) assisted downlink system. We first determine the user grouping with a normalised channel similarity metric so as to allow a large number of users with non-zero achievable private message rates. Through the private message removal (PMR) of these users, we aim to investigate the QoS vs. energy efficiency trade-off. Numerical results indicate a peak of ~$10\%$ increase in the network energy efficiency for the proposed normalised channel similarity metric based user grouping with scheduled PMR.
\end{abstract}

\section{Introduction}
    The iterative evolution of demands for future wireless networks roughly every ten years constantly probes the possibility of providing higher Quality-of-Service (QoS) for communication users \cite{Gen_6G}. With the constant surge of novel technologies in wireless communication systems, design techniques and signal processing facilities, future wireless networks are envisioned to meet these expectations albeit with as few network resources as possible \cite{Gen_6G}. 
    
    This quest for energy efficient resource allocation despite the growing demands on higher QoS, has placed rate splitting multiple access (RSMA) as a front runner among other multiple access (MA) schemes \cite{MA_Gen}. RSMA distinguishes itself by offering diverse advantages in system design such as high spectral and energy efficiency \cite{ener_effi, ener_effi2}, along with robustness against imperfect channel state information at the transmitter (CSIT) \cite{reifert2022comeback, RS_CSIT2, RS_CSIT3, RS_CSIT4}. 
    
    Within the general idea of splitting user messages into common and private parts, RSMA captures the essence of both treating interference as noise (TIN) with space-division multiple access (SDMA) and fully decoding the interference through successive interference cancellation (SIC) with non-orthogonal multiple access (NOMA) \cite{ener_effi3}. The authors in \cite{ener_effi3} discuss the abridging nature of rate-splitting (RS) as a tailored combination of these two extremes with a relatively lower computational complexity (subject to the number of SIC layers). 
    
    Further in \cite{ener_effi3}, RS-common message decoding (RS-CMD) was proposed for inter-user interference mitigation. Based on a heuristic segregation of users, common messages of other users from the same group are sequentially decode in \cite{ener_effi3}. These decoded messages are then successively cancelled at the communication receivers, contributing to better inter-user interference mitigation compared to other MA schemes. In \cite{RS_vs_NOMA, RS_THz_IRS}, the superiority of RSMA as a MA scheme in terms of achievable QoS and dexterity between the SDMA and NOMA extremes was demonstrated. 
    
    In this paper, we adopt a normalised channel similarity metric based user grouping for the RS-CMD downlink system, as opposed to the heuristic based segregation in \cite{ener_effi3}. We maximize the network energy efficiency for optimal resource allocation under minimum QoS constraints at the users for this RS-CMD downlink system. Additionally, we study the impact of scheduled private message removal (PMR) in the QoS vs. the network energy efficiency evaluation. 
    
    This qualitative comparison is motivated from the deliverable standard/high definition (SD/HD) trade-off for streaming service users over congested networks. For instance, with increase usage of bandwidth from residential areas during covid-19, it was a huge issue for streaming service providers to satisfy HD requirements for all users, forcing the switch to SD quality streams \cite{HD_SD_Apple}. In the context of RSMA, the achievable QoS with the split common and private messages could be regarded as SD and HD services, respectively. We employ an extended version of the channel similarity metric \cite{RS_grps} to determine users with inefficient private rates and downgrade their service to SD from HD through PMR in this paper. 
    
    An investigative study on the impact of scheduled private message removal (PMR) for these inefficient users on the network energy efficiency, and the overall power saved in the system is the objective of this paper. It is noteworthy to mention that once all the private messages are removed, the RS-CMD scheme in the system equivalently transitions into the TIN with SDMA scheme.

\section{System Model} \label{Sys_model}

\begin{figure}
        \centering
        \includegraphics[width= 0.77\columnwidth]{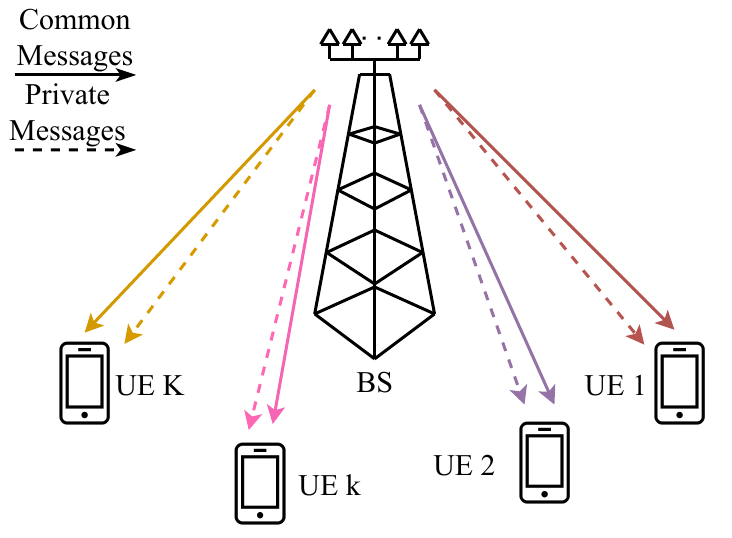}
        \caption{RSMA assisted BS with $N$ transmit antennas, serving $K$ single antenna users. Scheduled private message removal (PMR) performed on users with inefficient rates for Quality of service (QoS) vs. energy efficiency trade-off evaluation.}
        \label{System_Model}
    \end{figure}

In this paper, we consider the RSMA enabled downlink system depicted in Fig.~\ref{System_Model}. The network consists of a basestation (BS) equipped with $N \geq 1$ transmit antennas, serving $\mathcal{K} = \{1,2,\dots,K\}$ single antenna users. 
    
    We consider the RSMA scheme to satisfy the QoS requirements of SD and HD users represented by minimum data rate of ${R}^{\mathsf{SD}}$ and ${R}^{\mathsf{HD}}$, respectively. Without loss of generality, we assume perfect CSI of the users at the BS. 
    
    The transmit signal at the BS for user $k$, with $s_k^p$ and $s_k^c$ representing the intended private and common messages with unit variances, respectively, is expressed as
    \begin{equation} \label{Tx_sig}
        \mathbf{x}=\sum_{k\in\mathcal{K}^p} \mathbf{w}_{k}^p s_k^p+\sum_{k\in\mathcal{K}^c}\mathbf{w}_{k}^c s_k^c,
    \end{equation}
    where $\mathcal{K}^p \subseteq \mathcal{K}$ and $\mathcal{K}^c \subseteq \mathcal{K}$ represent the set of private and common message users respectively. Accordingly in \eqref{Tx_sig}, $\mathbf{w}_k^p \in \mathcal{C}^{N \times 1}$ and $\mathbf{w}_k^c \in \mathcal{C}^{N \times 1}$ are the private and common beamformers for user $k$, respectively. The transmitted private $s_k^p$ and common messages $s_k^c$ can be chosen from an independent and identically distributed (i.i.d) Gaussian codebook. Consequently, the total transmit power $P^{\mathsf{Tr}}(\mathbf{w})$ at the BS is constrained to 
    \begin{equation}  \label{Pow_const}
    \sum_{k\in\mathcal{K}}\left(\norm{\mathbf{w}_{k}^p}_2^2+\norm{\mathbf{w}_{k}^c}_2^2\right)\leq P^{\mathsf{Tr}}(\mathbf{w}).
    \end{equation} 
    Now, the channel matrix $\mathbf{H} = [\mathbf{h}_1,\dots, \mathbf{h}_K] \in \mathcal{C}^{N \times K}$, represents the communication channels between the $N$ transmit antennas at the BS and the $K$ communication users. The received signal at user $k\in\mathcal{K}$ through its channel $\mathbf{h}_k$ can be expressed as
    \begin{equation} \label{Rx_sig}
        y_k=\mathbf{h}_k^H\mathbf{x}+n_k, \;\; 
    \end{equation}
    where $n_k\sim\mathcal{N}(0,\sigma_k^2)$ is the zero-mean additive white Gaussian noise (AWGN) at user $k$, with noise variance $\sigma^2_k$. 
    We first determine the set of users $\mathcal{M}_k$ that decode the common message $s_k^c$ of user $k$ and the set of users $\mathcal{Z}_k$ whose common messages are decoded at user $k$ given by
    \begin{subequations} \label{Sim_metric}
    \begin{equation} \label{Sim_matrx}
        \mathcal{M}_k = \{j\in \mathcal{K}| \; \text{user $j$ decodes $s_k^c$ }\},
    \end{equation}
    \begin{equation} \label{Phi_k}
        \mathcal{Z}_k = \{j\in \mathcal{K}| k \in \mathcal{M}_j\}.
    \end{equation}
    \end{subequations}
    
    Using these definitions, the received signal \eqref{Rx_sig} at user $k$ can be rewritten in terms of desired and undesired parts as 
    \begin{align} \label{Rx_sig2}
        y_k &=\overbrace{\left(\mathbf{h}^H_k\mathbf{w}_k^ps_k^p+\sum_{j\in\mathcal{Z}_k}\mathbf{h}_k^H\mathbf{w}_j^cs_j^c\right)}^{\text{Desired signal}}+\nonumber \\ 
            &\underbrace{\sum_{j\in\mathcal{K}\backslash\{k\}}\mathbf{h}_k^H\mathbf{w}_j^ps_j^p+\sum_{l\in\mathcal{K}\backslash\{\mathcal{Z}_k\}}\mathbf{h}_k^H\mathbf{w}_l^cs_l^c+n_k}_{\text{Interference + noise}}.
    \end{align}   
    
    In \eqref{Rx_sig2}, we successively decode and cancel the common messages of users in $\mathcal{Z}_k$ \cite{ener_effi3}. The set $\mathcal{Z}_k$ determines the effectiveness of RS-CMD in mitigating the interference from streams of unintended groups. To this end, the set $\mathcal{Z}_k$ is constructed based on a channel similarity metric from \cite{RS_grps}.
    
    In the following subsection, we first motivate the necessity of private message removal (PMR) of certain users. Further, we discuss the ideal RS group formation required to demonstrate the pronounced impact of PMR. Based on this, we define the construction of the set $\mathcal{Z}_k$ using the channel similarity metric.
    
    \subsection{Channel similarity metric \& RS Group Formation} \label{Chan_sim_sec}
    Within the RS-CMD regime, we define inefficient users as the users that require unreasonably high transmit powers for rather low private message rates. We evaluate the PMR vs. the energy efficiency trade-off by removing all the private messages in the system. Since we also remove the corresponding beamforming power associated with the removed private rate, PMR of the inefficient users improves the energy efficiency.
    
    In this context, it is beneficial to form multiple smaller RS-CMD groups, instead of fewer large ones. This is because, utilizing smaller RS-CMD groups entails the benefit of more robustness with respect to (w.r.t) changes in rate and power allocation. In other words, sending out more common messages would result in more PMR events for energy efficiency evaluation. 
    
    To this end, we only allow every user to decode a maximum of $1$ alien common message and its own, before decoding its own private message. This entails that we set $|\mathcal{Z}_k| = D = 2$ as the number of decoding layers for each user $k$.
    
    In more details, we employ the channel similarity metric $\mathbf{R}$ for the RS-CMD group formation, given by
    \begin{align} \label{Sim_metr}
    [\mathbf{R}]_{k,j} = \frac{\left|\mathbf{h}_k^H\mathbf{h}_j\right| }{\norm{\mathbf{h}_k}\norm{\mathbf{h}_j}}.
    \end{align}
    The channel similarity metric $\mathbf{R}$ groups users with similar channel vectors, making it a channel alignment based grouping approach. The channel similarity between  any two channels would have the same value, i.e., $\mathbf{R}_{ij} = \mathbf{R}_{ji}$. The resulting matrix $\mathbf{R}$ in \eqref{Sim_metr}, hence, would have a symmetric structure. This symmetry introduces ambiguity when determining the owner of the common message for the formed groups. 
    
    We break this symmetry by employing a scaling parameter on $\mathbf{R}$ in the form of the normalised channel strengths $\tilde{\mathbf{h}}$, expressed as 
    \begin{align} \label{norm_chan}
        \tilde{\mathbf{h}} = \frac{1}{{\underset{k\in \mathcal{K}}{\max}{||\mathbf{h}_k||_2^2}}} \left[||\mathbf{h}_1||_2^2,\dots,||\mathbf{h}_K||_2^2\right]^T.
    \end{align}
    With \eqref{Sim_metr} and \eqref{norm_chan}, we formulate the extended channel similarity matrix $\Tilde{\mathbf{R}}$ as
    \begin{align} \label{Norm_Sim_met}
    \tilde{\mathbf{R}} =  \mathsf{diag}(\tilde{\mathbf{h}})\mathbf{R}.
    \end{align}
    
    Now, we determine the user $i_{\mathsf{max}}$ with the highest value in $\tilde{\mathbf{R}}$ for decoding the common message of user $j_{\mathsf{max}}$ as 
    \begin{equation} \label{Sim_matrx2}
        i_{\mathsf{max}}, j_{\mathsf{max}} = \arg \max_{i,j} (\tilde{\mathbf{R}}_{ij}).
    \end{equation}
    Afterwards, we set $[\tilde{\mathbf{R}}]_{i_{\mathsf{max}}, j_{\mathsf{max}}} = 0$ and the process is repeated until the decoding layer limit $|\mathcal{Z}_k| = D = 2, \forall \; k\in \mathcal{K}$ is reached. Within the formed RS groups, we first allow the user with the stronger channel to decode the common message of the weaker user. The user with the stronger channel then decodes it's own private message in the end after successive interference cancellation (SIC). Accordingly, the SINR of user $k$ decoding the common message of user $i$ can be expressed as
    \begin{equation} \label{SINR_C}
        \begin{split}
            \gamma^c_{k,i} &= \frac{\left|\mathbf{h}_k^H\mathbf{w}_i^c\right|^2}{T_k+\sum\limits_{l\in\mathcal{K}\backslash\{\mathcal{Z}_k\}}\left|\mathbf{h}_k^H\mathbf{w}_l^c\right|^2+\sum\limits_{m\in\mathcal{Z}_k\backslash i}\left|\mathbf{h}_k^H\mathbf{w}_m^c\right|^2},
        \end{split}
    \end{equation}
    where $T_k = \sum\limits_{j\in\mathcal{K}}\left|\mathbf{h}_k^H\mathbf{w}_j^p\right|^2+\sigma_k^2$. After SIC of the decoded common messages corresponding to $\mathcal{Z}_k$ in \eqref{SINR_C}, the private stream SINR of user $k$ is given by 
    \begin{equation} \label{SINR_P}
        \gamma_k^p =\frac{\left|\mathbf{h}_k^H\mathbf{w}_k^p\right|^2}{\sum\limits_{j\in\mathcal{K}\backslash k }\left|\mathbf{h}_k^H\mathbf{w}_j^p\right|^2+\sum\limits_{l\in\mathcal{K}\backslash\{\mathcal{Z}_k\}}\left|\mathbf{h}_k^H\mathbf{w}_l^c\right|^2+\sigma_k^2}.
    \end{equation}
    Let $\xi_k^c$, $\xi_k^p$ be the achievable common and private message rates at user $k$. With $B$ representing the transmission bandwidth, the achievable private and common rates, respectively, at user $k$ are upper bounded as
    \begin{align} \label{Rate_constr}
        B\log_2(1+\gamma_k^p) &\geq \xi_k^p \; \forall\; k\in\mathcal{K},    \\
        \min_{i\in\mathcal{M}_k}\{B\log_2(1+\gamma_{i,k}^c)\} &\geq \xi_k^c \; \forall\; k\in\mathcal{K}.
    \end{align}  	

\section{Energy Efficiency- QoS trade-off} \label{Opt} 
    Each of the formed RS-CMD groups now posses a user $j$ that decodes the common message of a certain user $k$ based on the normalised channel similarity metric, and user $k$ is responsible for decoding its own common message. As discussed in the earlier subsection, this also ensures that we have several potential candidates for PMR events. With our objective to study the trade-off between energy efficiency and QoS by providing only common messages (SD service) to inefficient users through PMR rather than both common and private messages (HD service), we then design the beamforming optimization problem for the RSMA downlink system. 
\subsection{Scheduling of PMR}
The central benchmark for our comparison is the case without any PMR event, i.e., all private messages are present. Once the energy efficiency is evaluated, we re-optimize the problem for each PMR event with the corresponding power from the removed private message being unavailable for transmission (for power saving purposes) at the BS. For instance, let $P^{Tr}$ be the total available transmit power before any PMR events, then the first PMR event over a certain user $l$ would result in the new available transmit power $P^{Tr'} = P^{Tr}- \norm{\mathbf{w}_l^p}^2_2$.

Each such PMR event needs to satisfy the following conditions:
Let $\mathcal{S}$ and $\mathcal{H}$ denote the sets of users that qualify and do not qualify for PMR, respectively, such that 
\begin{align} \label{HD_SD_Sets}
    \mathcal{H}\cup \mathcal{S} &= \mathcal{K} \\
    \mathcal{H}\cap \mathcal{S} &= \mathcal{\varnothing}
\end{align}
then the choice of each user, say user $e$, undergoing PMR is made based on the following conditions: 
\begin{itemize}
    \item user $e$ has a common rate allocated for decoding his own common message ($\xi_k^c>0$)
    \item user $e$ has a private rate allocated for decoding his private message ($\xi_k^p>0$)
    \item if there is a set of users $\mathcal{E}$ that fulfill these criteria, then the user with the smallest non-zero private rate ($\xi_k^p>0$) is chosen as user $e$
    \item if no such user can be found, PMR scheduling stops.
\end{itemize}
Consequently, the private beamformer power of user $e$ is made unavailable at the overall transmit power for next optimization as $P^{Tr'} \leftarrow P^{Tr} - \norm{\mathbf{w}_{e}^p}_2^2$, and user $e$ is removed from $\mathcal{H}$ to be added in $\mathcal{S}$ as 
\begin{align} \label{HD_SD_Sets}
    \mathcal{H}' &= \mathcal{H}\backslash \{e\} \\
    \mathcal{S}' &= \mathcal{S}\cup\{e\}.
\end{align} 
\subsection{Problem Formulation}
We update the sets $\mathcal{H}'$ and $\mathcal{S}'$ for re-optimizing over each PMR event accordingly using the problem formulated below. As stated in the previous sections, we perform energy efficiency maximisation subject to QoS constraints iteratively for each PMR event while updating the transmit power constraint using the formulation below:
\begin{subequations} \label{Opt_prob:P1}
\begin{align}
\!\underset{\{\mathbf{w}_k,\mathbf{t}_k,\boldsymbol{\xi}_k\}_{k\in \mathcal{K}}}{\text{maximize}} \qquad& \Upsilon =\frac{\sum_{k\in\mathcal{K}}\left(\xi_k^p+\xi_k^c\right)}{P^{\text{Tr}}(\mathbf{w})+P^{\text{Circ}}}\label{eq:optProb}\\
\text{subject to} \qquad &\xi_k^p-B\log_2(1+t_k^p)\leq0,\;\forall k\in\mathcal{K},\label{eq:constraint1}\\
               &\xi_k^c-B\log_2(1+t_k^c)\leq0,\;\forall k\in\mathcal{K},\label{eq:constraint2}\\
               &\mathbf{t}_k\geq0,\;\forall k\in\mathcal{K},\label{eq:constraint3}\\
               &\boldsymbol{\xi}_k\geq0,\;\forall k\in\mathcal{K},\label{eq:constraint4}\\
               &t_k^p\leq\gamma_k^p(\mathbf{w}),\;\forall k\in\mathcal{K},\label{eq:constraint5}\\
               &t_k^c\leq\gamma_{i,k}^c(\mathbf{w}),\;\forall k\in\mathcal{K},\;\forall  
                   i\in\mathcal{M}_k,\label{eq:constraint6}\\
              & |(\xi_k^p+\xi_k^c) - R^\mathsf{HD}| \leq 0,\; \forall k \in \mathcal{H},\label{eq:constraint7}\\
              &|(\xi_k^p+\xi_k^c) - R^\mathsf{SD}| \leq 0,\; \forall k \in \mathcal{S},\label{eq:constraint8}
\end{align}
\end{subequations}
where $P^{\text{Circ}}$ from \eqref{eq:optProb} representing the power required for network circuitry is treated as constant, constraints \eqref{eq:constraint1}, \eqref{eq:constraint2} represent the QoS requirements. The slack variables $\mathbf{t}_k = [t_k^p,t_k^c]$ in \eqref{eq:constraint3}, \eqref{eq:constraint5} and \eqref{eq:constraint6} are introduced to convexify the rate expressions. Additionally, we also incorporate the constraints $R^{\mathsf{HD}}$ for users in  $\mathcal{H}$ and $R^{\mathsf{SD}}$ for the users in $\mathcal{S}$ in \eqref{eq:constraint7}) and \eqref{eq:constraint7}, respectively, to ensure that the QoS requirements are strictly enforced. This way, the energy efficiency for the desired rates is improved by scaling down the beamformers. As a consequence, we prevent unnecessary power allocation to users with already high-quality channels and avoid wasted resources which would otherwise be spent on achieving unnecessarily high rates.

Further, the problem in \eqref{Opt_prob:P1} can be solved by employing successive convex approximation (SCA) in combination with the Dinkelbach transformation \cite{syn_ben}. However, the constraints in \eqref{eq:constraint5} and \eqref{eq:constraint6} need to be represented in an alternative form since they currently define a non-convex feasible set making the Dinkelbach approach inefficient. This can be overcome by first rewriting \eqref{eq:constraint5} and \eqref{eq:constraint6} in an equivalent convex form as illustrated in \cite{syn_ben}. The fractional objective function from \eqref{Opt_prob:P1} is solved using the Dinkelbach transformation with the Dinkelbach coefficient $\lambda$ \cite{dinkelba} defined as
\begin{align} \label{Dinkel}
    \lambda = \frac{\sum_{k\in\mathcal{K}}\left(\xi_k^p+\xi_k^c\right)}{P^{\text{Tr}}(\mathbf{w})+P^{\text{Circ}}},
\end{align}
where we feed the optimal values from the previous iteration in \eqref{Dinkel} to iteratively optimize the beamformers while maximizing the new objective function in \eqref{new_obj} until convergence. 
\begin{align} \label{new_obj}
    \Upsilon' = \sum_{k\in\mathcal{K}}\left(\xi_k^p+\xi_k^c\right) - \lambda(P^{\text{Tr}}(\mathbf{w})+P^{\text{Circ}})    
\end{align}
Finally, we determine the converged values for network energy efficiency using the formulations from above, while updating the beamformers $\mathbf{w}$ and the available transmit power $P^{\text{Tr}(\mathbf{w})}$ for each PMR event.

\section{Numerical Results} \label{Sim_results}    
    We consider the RS-CMD assisted downlink system spanning over an area of $[-250,250] \times [-250,250]\,~{\mathsf{m}^2}$, with the BS situated in the center and equipped with $N = 8$ transmit antennas, serving $K = 12$ single antenna users. We set $P^{\mathsf{Tr}} = \SI{35}{dBm}$ as the total available transmit power before the PMR events, $P^{\mathsf{Circ}} = \SI{37}{dBm}$, bandwidth $B = \SI{10}{MHz}$, noise power $\sigma^2$ at $\SI{-102}{dBm}$, achievable QoS requirements at $R^{\mathsf{HD}}= \SI{8}{Mbps}$ and $R^{\mathsf{SD}}= \SI{4}{Mbps}$.  
    
    We further set the number of decoding layers for the RS-CMD groups at $D = \left|\mathcal{Z}_k\right| = 2$ for any user $k$ as stated and justified in \ref{Chan_sim_sec}. We evaluate the system performance in terms of network energy efficiency and optimal achievable common rates w.r.t the PMR events. Among the $K = 12$~users in the RS-CMD beamforming optimization problem from \ref{Prob_form}, $8$~users were identified to have non-zero private and common rates as a result of \eqref{Opt_prob:P1}. This implies that we begin with $\left|\mathcal{H}\right| = 8$ before the PMR events based on the scheduling defined in \ref{PMR_Sched}.  
    
    It has to be noted that the concept of demoting users from HD to SD service applies only to these $8$~users with non-zero-private rates. We therefore have at most $8$ PMR events (and hence $8$~SD service users, not $12$).
    
    \begin{figure}
        \begin{center}
            \includegraphics[width = \columnwidth]{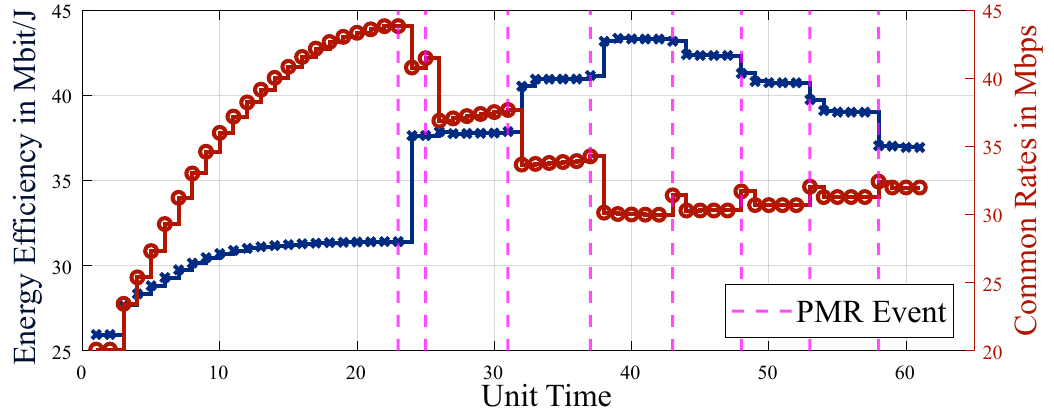}
        \end{center}
    \caption{Performance comparison w.r.t scheduled PMR events in terms of system energy efficiency $\Upsilon$ in $\mathsf{Mbit/J}$ ($\mathsf{-x}$ curves) and common rates in $\mathsf{Mbps}$ ($\mathsf{-o}$ curves).}
        \label{fig:CR_EE_Iter}
    \end{figure}
    In Fig.~\ref{fig:CR_EE_Iter}, the impact of PMR events can be observed on the energy efficiency and common rates plots on the left and right y-axes, respectively. The common rate curve can be seen to converge to a peak of about $\SI{43}{Mbps}$ before the first PMR event in Fig.~\ref{fig:CR_EE_Iter}. Conversely, the least recorded energy efficiency of the system at $\SI{31}{Mbit/J}$ can also be identified to be at this same point in Fig.~\ref{fig:CR_EE_Iter}. This observation justifies our intuitive idea of the RS-CMD system having inefficient private message users (and hence the low energy efficiency) whose QoS are met with the high common rates.
    
    With each PMR event in Fig.~\ref{fig:CR_EE_Iter}, the optimization problem from \eqref{Opt_prob:P1} is solved until convergence while the sets $\mathcal{H}$ and $\mathcal{S}$ are updated based on the definitions from \ref{PMR_Sched}. Further in Fig.~\ref{fig:CR_EE_Iter}, the common rates curve can be observed to drop from the first to the fourth PMR event. This indicates that each of the four PMR events are performed over the most inefficient private message user during that energy efficiency maximization. Further, harsh jumps in energy efficiency can be observed during this regime in Fig.~\ref{fig:CR_EE_Iter}. This is because, the reduction of the QoS demand ($R^{\mathsf{HD}}$ to $R^{\mathsf{SD}}$) for the affected user far outweighs the spent power for the private message in this regime. 
    
    Finally in Fig.~\ref{fig:CR_EE_Iter}, as PMR is performed over the last four remaining private message users, the common rates rise to meet the QoS requirements, at the expense of reduced energy efficiency. This indicates that all PMR events during this regime are ineffective and are being carried out over users with efficient private rates.
    \begin{figure}
        \begin{center}
            \includegraphics[width = \columnwidth]{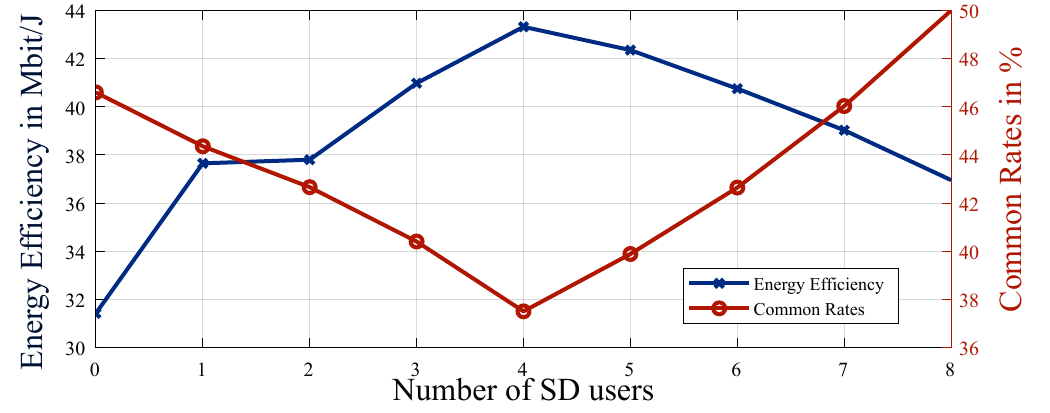}
        \end{center}    
        \caption{Energy efficiency $\Upsilon$ (in $\mathsf{Mbit/J}$) and common rates (in $\%$ w.r.t total rate) performance comparison vs. the number of SD users in the system.}
        \label{fig:CR_EE_SD}
    \end{figure}
     
    On the left y-axis of Fig.~\ref{fig:CR_EE_SD}, a comparison of the number of SD users in the network vs. the corresponding impact on energy efficiency is shown. The peak energy efficiency can be observed at $\SI{43}{Mbit/J}$, when the number of SD users~$=4$ in the network. This is because, by the end of this PMR regime, i.e. at number of SD users~$=4$, the remaining HD users in the network possess very efficient private rates. This further explains the plunge in the $\%$ of common rates curve to $38\%$ in Fig.~\ref{fig:CR_EE_SD} on the right y-axis, exactly when the energy efficiency reaches its peak. 
    
    While the energy efficiency curve follows a trend of steep increase and then a decrease with increasing SD users, a complementary reaction can be observed with the $\%$ of common rates curve in Fig.~\ref{fig:CR_EE_SD}. These observations made from Fig.~\ref{fig:CR_EE_SD} displaying the complementary behaviour of energy efficiency vs. $\%$ of common rates curve, also agree with the results from Fig.~\ref{fig:CR_EE_Iter}.
    \begin{figure}
        \begin{center}
            \includegraphics[width = \columnwidth]{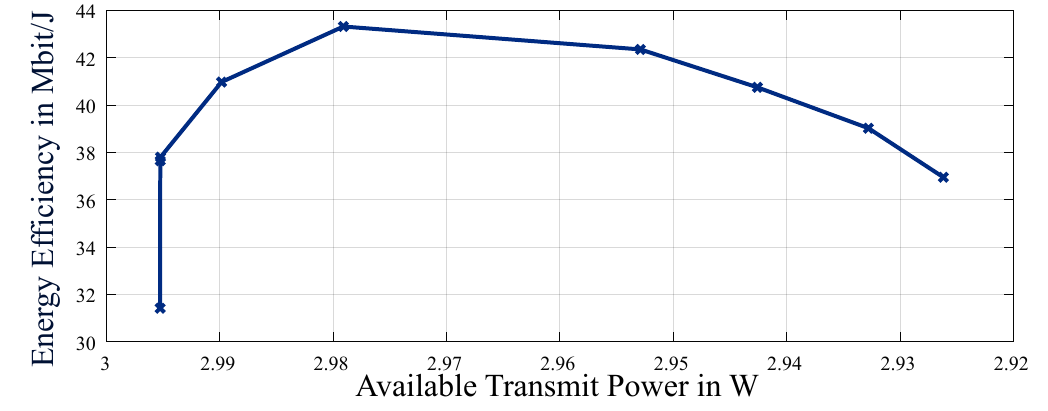}
        \end{center}
        \caption{A comparison of energy efficiency $\Upsilon$ (in $\mathsf{Mbit/J}$) vs. reduced transmit power (in $\mathsf{W}$) available for energy efficiency maximization after each PMR event.}
        \label{fig:EE_TransPow}
    \end{figure}
    \vspace{-1pt}
    
    Finally, Fig.~\ref{fig:EE_TransPow} shows a comparison of the network energy efficiency vs. the decreasing available transmit power over the PMR events. Each point on the energy efficiency curve corresponds to a PMR event in Fig.~\ref{fig:EE_TransPow}. As mentioned in \ref{PMR_Sched}, each PMR event results in the reduction of the available transmit power, corresponding to the removed private beamformer. The extreme gain in energy efficiency during the first PMR event with little reduction in the available transmit power indicates the inefficiency of the removed private user. By removing this private rate, the QoS requirement for the most inefficient user got reduced from $R^{\mathsf{HD}}$ to $R^{\mathsf{SD}}$, which is highly beneficial for the network energy efficiency. Fig.~\ref{fig:EE_TransPow} further shows a distinct maximum in the network energy efficiency at $\SI{43}{Mbit/J}$ for the case of having $4$~SD users, agreeing with the observations from Fig.~\ref{fig:CR_EE_Iter} and Fig.~\ref{fig:CR_EE_SD}. Finally, it is noteworthy to observe that the last PMR event in Fig.~\ref{fig:CR_EE_Iter}, Fig.~\ref{fig:CR_EE_SD} and Fig.~\ref{fig:EE_TransPow} corresponds to the case where all the private messages in the system are removed. This marks the transition of the RS-CMD scheme to the TIN with SDMA scheme.  

\section{Conclusion} \label{Conclusion}
    In this paper, we explore the RS-CMD assisted downlink system with the objective of energy efficiency maximization under minimum QoS constraints. Based on a normalised channel similarity metric and controlled RS group formation, we evaluate the impact of private message removal (PMR) of the users in the network. Through a minimal back-off in terms of QoS, PMR of inefficient users was shown to improve the energy efficiency. This intuitive idea of PMR of users with the smallest non-zero private rates and its impact on the network energy efficiency has been explored, and verified through numerical simulations. The potential of energy-aware RS-CMD to improve the user QoS has been investigated.

\vspace{12pt}	
\end{document}